\documentclass{autart}
\usepackage[english]{babel}
\usepackage{amsmath}
\usepackage{amsfonts}
\usepackage{amssymb}
\usepackage{graphicx}
\usepackage{epstopdf}

\newcommand{\Tr} {\mbox{\rm tr}}
\newcommand{\var} {\mbox{\rm var}}

\newcommand{\Var} {\mbox{\rm Var}\,}
\newcommand{\Cov} {\mbox{\rm Cov}\,}

\newcommand{\E}{{\mathbb E}\,}

\allowdisplaybreaks

\begin{document}

\begin{frontmatter}

\title{A new kernel-based approach to system identification\\with quantized output data} 

\thanks[footnoteinfo]{This work was  supported by the Swedish Research Council via the projects \emph{NewLEADS} (contract number: 2016-06079) and
\emph{System identification: Unleashing the algorithms} (contract number: 2015-05285),  by the MIUR FIRB project RBFR12M3AC - \emph{Learning meets time:
a new computational approach to learning in dynamic systems}, and by Progetto di Ateneo CPDA147754/14 - \emph{New statistical learning approach for multi-agents adaptive estimation and coverage control}. }

\author[First]{Giulio Bottegal},
\author[Second]{H\r akan Hjalmarsson},
\author[Third]{Gianluigi Pillonetto}

\address[First]{Department of Electrical Engineering, Eindhoven University of
Technology, Eindhoven, The Netherlands \\ (e-mail: g.bottegal@tue.nl)}
\address[Second]{Automatic Control Lab and ACCESS Linnaeus Centre, School of Electrical Engineering, KTH Royal Institute of Technology, Stockholm, Sweden (e-mail: hjalmars@kth.se)}
\address[Third]{Department of Information Engineering, University of Padova, Padova, Italy  (e-mail: giapi@dei.unipd.it)}


\begin{abstract}                          
In this paper we introduce a novel method for linear system identification with quantized output data. We model the impulse response as a zero-mean Gaussian process whose covariance (kernel) is given by the recently proposed stable spline kernel, which encodes information on regularity and exponential stability. This serves as a starting point to cast our system identification problem into a Bayesian framework. We employ Markov Chain Monte Carlo methods to provide an estimate of the system. In particular, we design two methods based on the so-called Gibbs sampler that allow also to estimate the kernel hyperparameters by marginal likelihood maximization via the expectation-maximization method. Numerical simulations show the effectiveness of the proposed scheme, as compared to the state-of-the-art kernel-based methods when these are employed in system identification with quantized data.
\end{abstract}

\end{frontmatter}

\section{Introduction}
\begin{figure}[ht]
\vskip 0.2in
\begin{center}
\centerline{\includegraphics[width=\columnwidth]{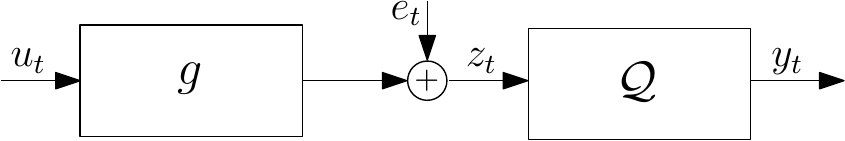}}
\caption{Block scheme of a system with quantized output data.}
\label{fig:block_scheme}
\end{center}
\vskip -0.2in
\end{figure}

Many applications in communications, control systems, bioinformatics, require modeling and prediction of dynamic systems with quantized output data (see e.g. \cite{bae2004gene}, \cite{carli2010quantized} and \cite{wang2010system}). In particular, in this paper we assume that an unknown input-output map is defined by the composition of a linear time-invariant dynamic system and a quantizer, as depicted in Figure \ref{fig:block_scheme}.
Estimation of this type of structure is a challenging task. Standard system identification techniques, such as least-squares or the Prediction Error Method (PEM) \cite{Ljung}, \cite{Soderstrom}, may give poor performances, because the presence of the quantizer can determine a significant loss of information on the system dynamics. Recently, there has been an increasing interest on developing new methods to solve this problem. Particular attention has been devoted to the case of binary measurements \cite{wang2003system}, \cite{wang2006joint}, also studying on-line recursive identification \cite{jafari2012convergence}, \cite{guo2013recursive}. In other works, e.g. \cite{colinet2010weighted}, the knowledge of a dithering signal is exploited to improve the identification performance. The problem of experiment design is analyzed in \cite{casini2011input}, \cite{casini2012input} and \cite{godoy2014novel}. More recently, system identification with non-binary quantizers has been also addressed in  \cite{marelli2013identification}, \cite{moschitta2015parametric}, \cite{guo2015asymptotically}, \cite{wang2015identification}. In \cite{godoy2011identification}, \cite{godoy2014identification}, \cite{chen2012impulse}, the system identification problem is posed as a maximum likelihood/a-posteriori problem. In particular, the framework proposed in \cite{chen2012impulse} uses the nonparametric kernel-based identification approach proposed in \cite{SS2010}, see also \cite{pillonetto2014kernel} for a survey.

Similarly to \cite{chen2012impulse}, in this paper we cast the problem of identifying a linear dynamic systems with quantized data in a Bayesian framework. To this end, we model the impulse response of the unknown system as a realization of a zero-mean Gaussian random process. The covariance matrix (in this context also called a \emph{kernel}) corresponds to the recently introduced \emph{stable spline kernel} (see \cite{SS2010}, \cite{SS2011}, \cite{bottegal2013regularized}). The structure of this type of kernel depends on two \emph{hyperparameters}, that need to be estimated from data. Following an empirical Bayes approach \cite{Maritz:1989}, the hyperparameters, together with the noise variance, are estimated via marginal likelihood maximization. These quantities are then used to compute the minimum mean-square estimate (MMSE) estimate of the impulse response. 

The whole procedure, which has been proven effective and relatively simple in the standard (non-quantized) setting, becomes more involved in the scenario under study. The MMSE system estimate requires the computation of an analytically intractable integral. To accomplish this task, in this paper we propose sampling methods based on Markov Chain Monte Carlo (MCMC) techniques. In particular, we design two different sampling techniques using the so called Gibbs sampler \cite{geman1984stochastic}, which often enjoys faster convergence properties compared to standard Metropolis-Hastings sampling techniques \cite{Gilks}. 

Another contribution of the paper is to show that the task of estimating the kernel hyperparameters (and the noise variance) can be accomplished using the proposed sampling techniques. Using the Expectation-Maximization (EM) method \cite{dempster1977maximum}, we design an iterative scheme for marginal likelihood maximization, where the E-step characterizing the EM method makes use of the Gibbs sampler (see also \cite{casella2001empirical}), and the M-step results in a sequence of straightforward optimization problems. Interestingly, the resulting estimation scheme can be seen as a generalization of the method proposed in \cite{godoy2014identification} to the Bayesian nonparametric framework and, differently from \cite{bottegal2015bayesian} and \cite{chen2012impulse}, it allows tuning all the kernel hyperparameters.

The paper is organized as follows. In the next section, we introduce the problem of identifying dynamic systems from quantized data. In Section \ref{sec:Bayesian_model},  we formulate the proposed Bayesian model and discuss its inference. In Section \ref{sec:sysid_method}, we describe the proposed identification method. Section \ref{sec:experiments} shows the results of several numerical experiments to assess the performance of the proposed method. Some conclusions end the paper.

\section{Problem formulation} \label{sec:problem_formulation}
We consider a linear time-invariant discrete-time dynamic system of the form
\begin{equation} \label{eq:sys1}
z_t = \sum_{i=1}^{+\infty} g_i u_{t-i} + e_t \,,
\end{equation}
where $\{g_t\}_{t=1}^{+\infty}$ is a bounded-input-bounded-output (BIBO) stable impulse response representing the dynamics of the system. We approximate the impulse response by considering the first $m$ samples only, namely $\{g_t\}_{t=1}^{m}$, where $m$ is assumed large enough to capture the system dynamics\footnote{All the results obtained in this paper can be also extended to the case $m=\infty$ but we prefer to consider a finite value for $m$
to simplify exposition. In addition, the assumption $m<N$, with $N$ the data set size, is typical of the system identification scenario.}.
The input $u_t$ is a measurable signal; for the sake of simplicity, we assume that the system is at rest until $t=0$, meaning $u_t = 0,\,t<0$. We shall not make any specific requirement on the input sequence (i.e., we do not assume any condition on persistent
excitation in the input \cite{Ljung}), requiring only $u_t \neq 0$ for some $t$. Notice, however, that even though the algorithm do not require any specifics of the input, the resulting estimate is of course highly dependent on the properties of the input \cite{Ljung}.
The output $z_t$ is corrupted by the additive zero-mean Gaussian white noise $e_t$ with variance $\sigma^2$, assumed unknown. Introducing the vectors 
$$
g := \begin{bmatrix} g_1 \\ \vdots \\ g_{m} \end{bmatrix} ,\,  \varphi_t^T := [ u_{t-1} \, \ldots \, u_{t-m} ] \,,
$$
an approximation of \eqref{eq:sys1} can be written in linear regression form, namely
\begin{equation} \label{eq:sys11}
z_t = \varphi_t^T g + e_t \,.
\end{equation}
The output $z_t$ is not directly measurable, only a quantized version being available, namely
\begin{equation}
y_t = \mathcal Q [z_t] \,,
\end{equation}
where $\mathcal Q$ is a known map of the type
\begin{equation}
\mathcal Q[x] = s_k \qquad \mbox{if } x \in (q_{k-1},\,q_k] \,,
\end{equation}
with $s_k \in \{s_1,\,\ldots,\,s_Q\}$ and $q_k \in  \{q_0,\,\ldots,\,q_Q\}$  (and typically $q_0 = -\infty$ and $q_Q = +\infty$).
\begin{rem}
A particular and well-studied case is the binary quantizer, defined as
\begin{equation}
\mathcal Q[x] = \left\{ \begin{array}{ll} -1 & \mbox{if } x < C \\
								  1 & \mbox{if } x \geq C
		\end{array}  \right. \,.
\end{equation}
It is well-known that a condition on the threshold to guarantee identifiability of the system is $C \neq 0$. In fact, when $C=0$, the system can be determined up to a scaling factor \cite{godoy2011identification}.
\end{rem}

We assume that $N$ input-output data samples $y_1,\,\ldots,\,y_{N}$, $u_0,\,\ldots,\,u_{N-1}$ are collected during an experiment. The problem under study is to estimate  $\{g_t\}_{t=1}^{m}$ using the collected measurements.

It is also useful to write the dynamics in the following vector notation
\begin{equation} \label{eq:sys2}
z = Ug + e\,,
\end{equation}
where $z$ and $e$ are $N$-dimensional vectors collecting the samples of $z_t$ and $e_t$, respectively, and $U \in \mathbb{R}^{N \times m}$ is a matrix whose $t$-th row corresponds to $\varphi_t^T$. Similarly, we denote by $y:= [ y_1 \,\, \ldots \,\, y_N ]^T$  the vector collecting the available quantized measurements.

\section{Bayesian modeling and inference} \label{sec:Bayesian_model}
\subsection{Impulse response prior}

In this paper we cast the system identification problem into a Bayesian framework, setting an appropriate prior on $g$. Following a Gaussian regression approach \cite{Rasmussen}, we model the impulse response as a zero-mean Gaussian random vector, i.e. we assume the following prior distribution
\begin{align}\label{eq:model_gh}
p(g;\,\lambda,\,\beta) \sim \mathcal N (0,\,\lambda K_\beta) \,. 
\end{align}
Here, $K_\beta$ is a covariance matrix whose structure depends on the \emph{shaping hyperparameter} $\beta$, and $\lambda \geq 0$ is a scaling factor.
In this context, $K_\beta$ is usually called a {\it kernel}  and determines properties of the realizations of $g$.
In particular, we choose $K_\beta$ from the family of \emph{stable spline kernels} \cite{SS2010}, \cite{SS2011}. Such kernels are specifically designed for system identification purposes and give clear advantages compared to other standard kernels \cite{bottegal2013regularized}, \cite{SS2010} (e.g. the Gaussian kernel or the Laplacian kernel, see \cite{Scholkopf01b}).
Motivated by its maximum entropy properties \cite{chen2016maximum}, in this paper we make use of the \emph{first-order stable spline kernel} (or \emph{TC kernel} in \cite{ChenOL12}). It is defined as
\begin{equation} \label{eq:ssk1}
\{K_\beta\}_{i,j} := \beta^{ \max(i,j)} \quad,\, 0 \leq \beta  <1 \,,
\end{equation}
with $\beta$ regulating the impulse response decay rate.

\subsection{Bayesian inference of systems with quantized output data} \label{sec:Gibbs}
In this section we describe the complete Bayesian model that stems from the prior adopted for the impulse response. Our main aim is to devise an MCMC-based sampling mechanism for inferring the model. We introduce the vector of  parameters $\eta = [\lambda \,\, \beta \,\, \sigma^2]$, seen as a deterministic quantity. 
\begin{figure}[!ht]
\begin{center}
{\includegraphics[width=6cm]{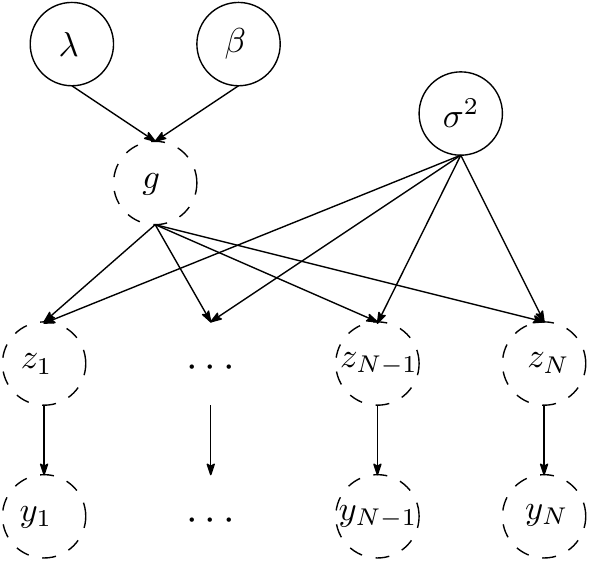}}
 \caption{\emph{Bayesian network describing the system model. Non-dashed nodes  denote deterministic or given quantities; dashed nodes denote random variables.}} \label{fig:bayesian_network}
\end{center}
\end{figure}
Figure \ref{fig:bayesian_network} depicts the Bayesian network describing the system identification problem with quantized data. The kernel hyperparameters $\lambda$ and $\beta$ determine the stochastic model of $g$, which, together with $\sigma^2$, in turn determines the distribution of the $z_t$. The output measurements $y_t$ are determined directly by the non-quantized outputs. Any kind of inference on this network crucially depends on the capability of performing Bayesian inference on the hidden nodes $g,\,\{z_t\}_{t=1}^N$, by computing functionals of the posterior density 
\begin{equation} \label{eq:target_density}
 p(g,\,z|y;\,\eta) 
\end{equation}
of the form
\begin{equation} \label{eq:functional_generic}
\int f(g,z) p(g,\,z|y;\,\eta) dg\,dz \,,
\end{equation}
where $f(g,z)$ is a general function of $g$ and $z$. Special cases of \eqref{eq:functional_generic} that will be used later comprise the MMSE of $g$ given $y$ and the computation of the marginal likelihood of $\eta$.

In the remainder of the section, we assume that $\eta$ is fixed.

\subsection{Method 1: sampling from the joint posterior} \label{sec:sampling1}
Quite unfortunately, analytical computation of \eqref{eq:functional_generic} is intractable, due to the involved structure of the posterior density. Instead, we can resort to Markov Chain Monte Carlo methods, i.e. use the approximation
\begin{equation} \label{eq:MCMC_approx}
\int f(g,z) p(g,\,z|y;\,\eta) dg\,dz \simeq \frac{1}{M} \sum_{k=1}^M f(g^{(k)},\,z^{(k)})\,,
\end{equation}
where $g^{(k)}$, $z^{(k)}$ are random samples drawn from \eqref{eq:target_density} and $M$ is a large integer. 
The problem is how to design an effective technique to sample from the distribution \eqref{eq:target_density}. If all the conditional probability densities of such a distribution are available in closed-form, the problem of sampling can be solved efficiently using a special case of the Metropolis-Hastings method, namely the Gibbs sampler (see e.g. \cite{Gilks}).
The basic idea is that each conditional random variable is the state of a Markov chain; then, by drawing samples from the conditional probability densities iteratively,
we converge to the stationary state of this Markov chain and generate samples of the desired distribution. The conditionals of \eqref{eq:target_density} are as follows.

\begin{enumerate}
\item $p(z_i|g,\,\{z_j\}_{j \neq i},\,y;\,\eta),\,i=1,\,\ldots,\,N$. First note that from \eqref{eq:sys2}, given $g$, $z_i$ is independent of $\{z_j\}_{j \neq i}$ and $\{y_j\}_{j \neq i}$. Therefore we have
\begin{equation}
p(z_i|g,\,\{z_j\}_{j \neq i},\,y;\,\eta)  = p(z_i|g,\,y_i;\,\eta) \,. 
\end{equation}
Now, using again \eqref{eq:sys2} we observe that, if $y_i$ were not given, then
\begin{equation} \label{eq:pz_gaussian}
p(z_i|g;\,\eta) \sim \mathcal{N}( \varphi_i^T g ,\,\sigma^2) \,.
\end{equation}
Knowing the quantized output $y_i$ permits to narrow the range of possible values of $z_i$. In particular, if $y_i = s_k$, for ant $k = 1,\,\ldots,\,Q$, we have
\begin{equation} \label{eq:truncated}
p(z_i|g,\,y_i = s_k;\,\eta)  \sim \mathcal{N}_{q_{k-1}}^{q_k} ( \varphi_i^T g,\,\sigma^2)\,,
\end{equation}
where $\mathcal N_{a}^{b}(\mu,\,\sigma^2)$ denotes a Gaussian distribution truncated below $a$ and above $b$, whose original mean and variance are $\mu$ and $\sigma^2$ respectively.

\item $p(g|z,\,y;\,\eta)$. Given $z$, information carried by $y$ becomes redundant and can be discarded. Due to the assumption on the distribution of the noise $e$, the vectors $g$ and $z$ are jointly Gaussian, so that the posterior density of $g$ given $z$ is also Gaussian. Combining the linear model \eqref{eq:sys2} with the prior \eqref{eq:model_gh} it is straightforward to obtain (see e.g. \cite{Anderson:1979})
\begin{equation} \label{eq:post_g}
p(g|z;\,\eta) \sim \mathcal{N}(m_g,\,P_g) \,,
\end{equation}
with
\begin{align} 
P_g & = \left( \frac{1}{\sigma^2}  U^T U + \left( \lambda_g K_\beta \right)^{-1}\right)^{-1} \label{eq:cov_g} \\
m_g & = \frac{1}{\sigma^2} P_g  U^T z = H z \label{eq:mean_g} \,,
\end{align}
with obvious definition of $H$.
\end{enumerate}
Algorithm 1 summarizes the computation of \eqref{eq:functional_generic} using the sampling mechanism  described in this subsection.
\begin{algorithm}[h!] 
\textbf{Algorithm 1}: Method 1 for inference \vspace{0.1cm}\\
Input: $\{y_t\}_{t=1}^N,\,\{u_t\}_{t=0}^{N-1},\,\eta$ \vspace{0.1cm} \\
Output: $\E[f(g,z)|y]$ 

Initialization: Compute initial value $g^{(0)}$ \\
For $k=1$ to $M+M_0$:
            \begin{enumerate}
                \item  Sequentially draw the samples $z_i^{(k)}$, $i=1,\,\ldots,\,N$, from $p(z_i|g^{(k-1)},\,y_i;\,\eta)$  %
                \item Draw the sample $g^{(k)}$ from $p(g|z^{(k)};\,\eta)$
               \end{enumerate}
Compute $\frac{1}{M} \sum_{k=M_0}^{M+M_0} f(g^{(k)},z^{(k)})$
\end{algorithm}

In Algorithm 1, the parameter $M_0$ is introduced. It represents the number of initial samples to be discarded and is also known as \emph{burn-in} \cite{meyn2009markov}. In fact, the conditional densities from which those samples are drawn are to be considered as non-stationary, because the associated Markov Chain takes a certain number of iterations to converge to a stationary distribution.

\subsection{Method 2: sampling from the marginalized posterior of $z$} \label{sec:sampling2}
Method 1 for sampling allows for computing expected values (with respect to the posterior $p(g,\,z|y;\,\eta)$) of any kind of function $f(g,\,z)$. 
For the problem under study however, we are particularly interested in computing the expected value (conditional on $y$) of $g$, and of quadratic functions in  $g$ and $z$, i.e.
\begin{equation} \label{eq:quadr_f}
f(g,z) = [g^T\,\,z^T] \begin{bmatrix}
A & B^T \\ B & C
\end{bmatrix} \begin{bmatrix} g \\ z \end{bmatrix} \,,
\end{equation}
for given $A \in \mathbb R^{N \times N}$, $B \in \mathbb R^{m \times N}$, $C \in \mathbb R^{m \times m}$. These type of functions will play a central role in  estimating the vector of parameters $\eta$ (see Section \ref{sec:sysid_method}).

The sampling method proposed in this subsection relies on the following result.
\begin{lem} \label{th:method2}
Let $f(g,\,z)$ be a quadratic function of the type \eqref{eq:quadr_f}. Then
\begin{align} \label{eq:th_result}
\E [f(g,\,z)|y]& = \Tr\{A P_g\}  \\ & + \int z^T(H^TAH + 2BH + C)z \, p(z|y;\eta)  dz \,, \nonumber
\end{align}
where $P_g$ and $H$ are defined in \eqref{eq:cov_g} and \eqref{eq:mean_g}.
\end{lem}
\paragraph*{Proof:}
We have
\begin{align}
& \int f(g,\,z) p(g,\,z|y;\,\eta) \,dg\,dz = \nonumber \\
& \int f(g,\,z) p(g|z;\,\eta) p(z|y;\,\eta) \,dg\,dz = \nonumber \\
& \int \left( \int f(g,\,z) p(g|z;\,\eta)\,dg \right) p(z|y;\,\eta) \,dz \,.  
\end{align}
We now focus on the internal integral in the above expression. Developing $f(g,\,z)$ and recalling the first and second moments of $p(g|z;\,\eta)$, given in \eqref{eq:post_g}, we obtain the following terms:
\begin{align*}
\int g^TAg \, p(g|z;\,\eta)dg &= \Tr\{AP_g\} +  m_g^TAm_g \\ & = \Tr\{AP_g\} + z^TH^TAHz \,, \\
\int z^TBg \, p(g|z;\,\eta)dg& = z^T B m_g = z^T B H z \,,  \\
\int z^TCz \, p(g|z;\,\eta)dg & = z^T C z \,.
\end{align*}
Combining the three terms yields \eqref{eq:th_result}. \hfill $\blacksquare$

A analogous result holds for the computation of $\E[g|y]$.
\begin{lem} \label{th:method2_2}
It holds that
\begin{align} \label{eq:th_result_2}
\E [g|y]& = H \int z \, p(z|y;\eta) \, dz \,, 
\end{align}
where $H$ is defined in \eqref{eq:mean_g}.
\end{lem}
Lemmas \eqref{th:method2} and \eqref{th:method2_2} state that the posterior expectations of $g$ and of quadratic functions in  $g$ and $z$, are equivalent to the expectation of specific functions with respect to the marginalized posterior of $z$ given $y$. This distribution corresponds to a multivariate truncated Gaussian; so, analytical computation of the relative integrals is quite challenging. Computing the integrals by sampling from $p(z|y;\,\eta)$ is instead a viable approach. 

An effective technique for sampling from $p(z|y;\eta)$ is again based on the Gibbs sampler. The idea is to iteratively draw samples from the conditional densities $p(z_i|\{z_j\}_{j \neq i},\,y_i;\,\eta)$. Generating samples from these distributions is relatively simple, because they are scalar truncated normal distributions. To see this, first consider the marginal distribution $p(z;\,\eta)$, which is a multivariate normal random vector with zero-mean and covariance matrix $\Sigma_z = \lambda U K_\beta U^T + \sigma^2 I$. 
Then 
\begin{equation}
p(z_i|\{z_j\}_{j \neq i};\,\eta) \sim N(m_{z_i},\,P_{z_i})\,,
\end{equation}
where 
\begin{align}
m_{z_i} & =  \Cov[z_i,\,\{z_j\}_{j \neq i}] \Var[\{z_j\}_{j \neq i}]^{-1} \label{eq:post_mean_z}\\
& \qquad\qquad \times [z_1\,\ldots z_{i-1}\,z_{i+1}\,\ldots\,z_N]^T \,, \nonumber \\
P_{z_i}& = \var[z_i] - \Cov[z_i,\,\{z_j\}_{j \neq i}] \Var[\{z_j\}_{j \neq i}]^{-1} \label{eq:post_cov_z} \\
& \qquad \qquad  \qquad \times \Cov[z_i,\,\{z_j\}_{j \neq i}]^T \nonumber \,.
\end{align}
Therefore, $p(z_i|\{z_j\}_{j \neq i},\,y_i;\,\eta) \sim N_{q_{k-1}}^{q_k}(m_{z_i},\,P_{z_i})$.

Algorithm 2 summarizes the sampling procedure  described in this subsection, for quadratic functions. The same algorithm can be used for computing the posterior expectation of $g$.
\begin{algorithm}[h!] 
\textbf{Algorithm 2}: Method 2 for inference \vspace{0.1cm}\\
Input: $\{y_t\}_{t=1}^N,\,\{u_t\}_{t=0}^{N-1},\,\eta$ \vspace{0.1cm} \\
Output: $\E[f(g,z)|y]$ 

Initialization: Compute initial value $g^{(0)}$ \\
For $k=1$ to $M+M_0$:
            \begin{enumerate}
                \item  Sequentially draw the samples $z_i^{(k)}$, $i=1,\,\ldots,\,N$, from $p(z_i|\{z_j\}_{j \neq i},\,y_i;\,\eta)$  %
            \end{enumerate}
Compute $\Tr\{AP_g\} + \frac{1}{M} \sum_{k=M_0}^{M+M_0} z^{(k)T} (H^TAH + 2BH + C)z^{(k)}$
\end{algorithm}
\begin{rem} \label{rem:sampling}
The quantities required to compute \eqref{eq:post_mean_z} and \eqref{eq:post_cov_z} can be retrieved from $\Sigma_z^{-1}$. Assume we are interested in generating a sample of $z_1$. To this end, partition $\Sigma_z^{-1}$ as follows
$$
\Sigma_z^{-1} = \begin{bmatrix}
s_1 & s_{j1}^T \\ s_{j1} & S_j
\end{bmatrix}\,,
$$
where $s_1 \in \mathbb{R},\,s_{j1} \in \mathbb{R}^{N-1},\,S_{j} \in \mathbb{R}^{N-1 \times N-1}$. Then $P_{z_1} = 1/s_1$ and $m_{z_1} = -\frac{s_{j1}^T}{s_1}[z_2\,\ldots\, z_{N}]^T$ (see e.g. \cite{noble1988applied}). This procedure turns out computationally convenient, because it is required to invert $\Sigma_z$ only one time, instead of computing the inversion of $N$ matrices of size $N-1 \times N-1$ (which would be necessary for computing $\Var[\{z_j\}_{j \neq i}]^{-1}$ for any $i = 1,\,\ldots,\,N$).
\end{rem}

\section{System identification from quantized output data} \label{sec:sysid_method}
In the previous section we have defined an adequate Bayesian framework to describe the problem of identification of systems from quantized output data. We have also presented two methods for inference of functions of $z$ and $g$. 
In this section, we describe the proposed system identification procedure. It relies on the so called empirical Bayes approach \cite{Maritz:1989} and consists of the following two steps:
\begin{enumerate}
\item Estimate the parameter vector $\eta$ via marginal likelihood maximization;
\item Compute the MMSE estimate of the impulse response, using the estimated parameter vector $\hat \eta$.
\end{enumerate}
In the following we analyze in detail the two steps composing the proposed system identification scheme.

\subsection{Marginal likelihood estimation of the parameter vector $\eta$}

In the empirical Bayes approach, the estimation of $\eta$ is tackled via marginal likelihood maximization, i.e., by computing
\begin{align} 
\hat \eta & = \arg \max_{\eta}  p(y;\,\eta) \label{eq:marg_lik} \\
& = \arg \max_{\eta} \int p(y,\,g,\,z;\,\eta)\, dg\,dz \,. \nonumber
\end{align}
In the standard (non-quantized) scenario, solving \eqref{eq:marg_lik} is straightforward because the marginal likelihood admits a closed-form expression \cite{pillonetto2014kernel}. This does not hold in the problem under study. Therefore, to solve \eqref{eq:marg_lik} we propose an iterative solution scheme based on the EM method, where we treat $g$ and $z$ as \emph{latent variables} that are iteratively estimated together with the parameter vector $\eta$. Let $\eta^{(n)}$ be the parameter estimate obtained at the $n$-th iteration of the EM method. Then, $n+1$-th update is obtained with the following steps:
\begin{description}
\item[(E-step)] Compute
    \begin{equation} \label{eq:estep}
    Q(\eta,\,\hat \eta^{(n)}) := \E \left[\log p(y,\,g,\,z;\,\eta) \right] \,,
    \end{equation}
    where the expectation is taken with respect to the posterior density $p(g,\,z|y;\,\hat \eta^{(n)})$, with $\eta$  fixed at the value $\eta^{(n)}$;
\item[(M-step)] Compute
    \begin{equation}
    \hat \eta^{(n+1)} = \arg \max_{\eta}  Q(\eta,\,\hat \eta^{(n)}) \,.
    \end{equation}
\end{description}
We first focus on performing the E-step. It requires the computation of \eqref{eq:estep}, which corresponds to the integral
\begin{equation} \label{eq:compl_lik_integral}
\int \log p(y,\,g,\, z;\,\eta) p(g,\,z|y;\,\hat \eta^{(n)}) \,dg\,dz \,. 
\end{equation}
\begin{prop} \label{th:complete_likelihood}
The complete likelihood admits the decomposition
\begin{align}
-2 \log p(y,\,g,\,z;\,\eta) & = \frac{1}{\sigma^2} f_1(g,\,z) + \frac{1}{\lambda} f_2(g,\,z,\,\beta) \\ 
& \!\!\!\!\!\!\!\!\!\!\!\!\!\!\!\!\!\!\!\!\!\!\!\!\!\!\!\!\!\!\!  + N \log \sigma^2 + \log \det \lambda K_\beta + \log p(y|z,\,g;\,\eta) \nonumber \,,
\end{align}
where $f_1(g,\,z)$ is a quadratic function of the type \eqref{eq:quadr_f}, with
$A = U^TU,\,B = U,\,C = I$, and $f_2(g,\,z,\,\beta)$ is also a quadratic function in $g$ and $z$, with  $A = K_\beta^{-1},\,B = 0,\,C = 0$.
\end{prop}
\paragraph*{Proof}
Using Bayes' rule we can decompose the complete likelihood as follows
\begin{align}
 \log p(y,\,g,\,z;\,\eta)&= \log p(y|z,\,g;\,\eta)  \\ & \quad+  \log p(z|g;\,\eta)  + \log p(g;\,\eta)  \,.\nonumber
\end{align}
The term $p(z|g;\,\eta)$ is a vectorized version of \eqref{eq:pz_gaussian}, so that
\begin{align} \label{eq:log_pz}
\log p(z|g;\,\eta)  =  -\frac{N}{2} \log \sigma^2 - \frac{1}{2\sigma^2} \|z - U g\|_2^2 \,.
\end{align}
The last term on the right hand side gives $f_1(z,\,g)$. Similarly, 
\begin{equation}
\log p(g;\,\eta) = \log \det \lambda K_\beta + g^T(\lambda K_\beta)^{-1}g \,,
\end{equation}
where the last term corresponds to $f_2(z,\,g,\,\beta)$. \hfill $\blacksquare$

The proposition reveals the nature of the complete likelihood, which is the summation of terms that are either constant or quadratic functions of $z$ and $g$, plus the term $\log p(y|z,\,g;\,\eta)$. As for this term, we note that $p(y|z,\,g;\,\eta)$ factorizes, each factor being of the type
\begin{equation} \label{eq:py_factor}
p(y_i|z,\,g;\,\eta)\! =\! \left\{ \begin{array}{ll}
	\! 1 & \mbox{if } y_i \! = \! s_k  \mbox{ and } z_i \! \in \! (q_{k-1},\,q_k] \\
	\! 0 & \mbox{otherwise} \end{array} \right.\! .
\end{equation}
When computing the integral of this term using the sampling mechanisms introduced in Section \ref{sec:Bayesian_model}, it is ensured that all the generated samples $z_i^{(k)}$ belong to the interval corresponding to the observed quantized value $y_i$. Hence, when we compute the expectation of $p(y|g,\,z;\,\eta)$ techniques of Section \ref{sec:Bayesian_model}, it is ensured that each factor \eqref{eq:py_factor} is always equal to 1 and thus $\log p(y|g,\,z;\,\eta) = 0$. 
Therefore, computing \eqref{eq:compl_lik_integral} reduces to computing the expectation of two quadratic functions, and this can be done using both the sampling mechanisms introduced in Section \ref{sec:Bayesian_model}. We denote by $\hat f^{(n)}_1$ and $\hat f^{(n)}_2(\beta)$, respectively, the expected value of $f_1(g,\,z)$ and $f_2(g,\,z,\,\beta)$ at the $n$-th iteration of the EM method (i.e., when the  $\eta^{(n)}$ of the parameter vector is available). Neglecting constant terms, we have
\begin{align} 
-2Q(\eta,\,\hat \eta^{(n)}) & = \frac{1}{\sigma^2} \hat f^{(n)}_1 + \frac{1}{\lambda}\hat f^{(n)}_2(\beta) \\
& \quad + N \log \sigma^2 + \log \det \lambda K_\beta \,. \nonumber
\end{align}

\begin{prop} \label{th:m-step}
Let
\begin{equation}
h(\beta) := m \log \hat f^{(n)}_2(\beta) + \log \det K_\beta\,.
\end{equation}
Then the EM update $\eta^{(n+1)} = [\lambda^{(n+1)}\, \beta^{(n+1)}\, \sigma^{2(n+1)}]$
is obtained as follows:
\begin{align}
\beta^{(n+1)} & = \arg \min_{\beta \in (0,\,1]} h(\beta)\,, \label{eq:betag_update}\\
\lambda^{(n+1)}& = m^{-1} \hat f^{(n)}_2(\beta^{(n+1)})   \label{eq:lambdag_update}\,, \\
\sigma^{2(n+1)}& = N^{-1} \hat f^{(n)}_1 \,. \label{eq:sigma_update}
\end{align}
\end{prop}
\paragraph*{Proof} 
Differentiating $-2 Q(\eta,\,\eta^{(n)})$ with respect to $\lambda$, one obtains that the minimum, as a function of $\beta$, is achieved at $\lambda^* =  m^{-1} \hat f^{(n)}_2(\beta)$. Inserting this value into $-2 Q(\eta,\,\eta^{(n)})$ yields $h(\beta)$, so that \eqref{eq:betag_update} and \eqref{eq:lambdag_update} follow. Finally, it is straightforward to see that \eqref{eq:sigma_update} is the minimizer of $-2 Q(\eta,\,\eta^{(n)})$. \hfill $\blacksquare$

The M-step results in a series of computationally attractive operations. The kernel scaling hyperparameter $\lambda$ and the noise variance $\sigma^2$ admit a solution in closed form. The shaping hyperparameter $\beta$ is updated solving a simple scalar optimization problem. This problem can be further simplified by using a factorization of the first-order stable spline kernel, see \cite{bottegal2016robust} for details.

\subsection{MMSE estimate of the impulse response}
Having an estimate $\hat \eta$ available, the MMSE estimate of $g$ corresponds to
\begin{align}
\hat g & := \E[g|y]  = \int g\, p(g|y;\,\hat \eta)\, dg \label{eq:int_1}\\
& = \int g\, p(g,\,z|y;\,\hat \eta) \,dg\,dz \,. \label{eq:int_2}
\end{align}
Again, this quantity can be computed using both the sampling methods described in the previous section.

We summarize the overall procedure for system identification from quantized data in the following algorithm.

\begin{algorithm}[h!] 
\textbf{Algorithm 3}: System identification with quantized output measurements \vspace{0.1cm}\\
Input: $\{y_t\}_{t=1}^N,\,\{u_t\}_{t=0}^{N-1}$ \vspace{0.1cm} \\
Output: $\{\hat{g_t}\}_{t=1}^m$

Initialization: Set an initial value of $\hat \eta^{(0)}$ \\
Repeat until convergence:
            \begin{enumerate}
                \item E-step: Compute $Q(\eta,\,\eta^{(n)})$ using Algorithm 1 or Algorithm 2   %
                \item M-step: update $\hat \eta^{(n+1)}$ according to Proposition \ref{th:m-step}
             \end{enumerate}
Compute $\hat g$ using  Algorithm 1 or Algorithm 2
\end{algorithm}
The initial value $\hat \eta^{(0)}$ can be set either randomly or by using values of the kernel hyperparameters and noise variance returned by the standard nonparametric method for non-quantized data (see \cite{SS2010}).

\subsection{Which sampling method?}
Algorithm 3 for quantized system identification works with both the sampling methods presented in Section \ref{sec:Bayesian_model}, because it requires the computation of the posterior expectation of quadratic functions of the type \eqref{eq:quadr_f} and of $g$. Choosing the sampling method depends on the user requirements. Method 1 allows for any type of inference and therefore it can be used to compute useful statistics such as confidence intervals on the estimate of $g$ and quantile estimation. On the other hand, Method 2 requires sampling only from $p(z|y,\,\eta)$; if implemented properly (see Remark \ref{rem:sampling}), this method is expected to have a faster convergence to the required expectations. The experiments presented in the next section show that the two sampling methods substantially give the same performance in terms of accuracy in reconstructing the true impulse response.

\section{Numerical experiments} \label{sec:experiments}

We test the proposed technique by means of 2 Monte Carlo experiments of 100 runs each. For each Monte Carlo run, we generate an impulse response by picking 10 pairs of complex conjugate zeros with magnitude randomly chosen in $[0,\,0.99]$ and random phase and, similarly, 10 pairs of complex conjugate poles with magnitude randomly chosen in $[0,\,0.92]$ and random phase. The obtained impulse response is rescaled in order to have a random $\ell_2$-gain in the interval $[2,\,4]$.  The goal is to estimate  $m=50$ samples of this impulse response from $N$ input-output data. The input sequences are realizations of white noise with unit variance. We compare the following estimators.
\begin{itemize}
\item \textbf{KB-GS-1}: This is the method described in this paper, making use of the sampling technique of Section \ref{sec:sampling1} (Algorithm 1). The parameter $M$, denoting the number of samples generated by the sampler, is set to $100$ for each iteration of the EM method, and to $500$ for the final computation of $g$ (last step of Algorithm 3). The burn-in phase consists of $M_0 = 100$ samples. Convergence of the EM method is established by a threshold on the relative difference of the current and previous parameter estimates, i.e. the method stops if the condition
$$
\frac{\|\eta^{(n+1)} -\eta^{(n)}\|_2^2}{\|\eta^{(n)}\|_2^2} \leq 10^{-3}
$$
is satisfied.  In our simulations, we have verified that our choice of $M$ was adequate to guarantee a good degree of approximation (the reader interested in convergence diagnostics is referred to, e.g., \cite[Ch. 11.4]{gelman2014bayesian}).
\item \textbf{KB-GS-2}: This is the method described in this paper, making use of the sampling technique of Section \ref{sec:sampling2} (Algorithm 2). The algorithm settings are the same as the previous method.
\item \textbf{KB-St.}: This is the standard nonparametric kernel-based method proposed in \cite{SS2010} and revisited in \cite{ChenOL12}. It is not designed for handling quantized data. The kernel adopted for identification is \eqref{eq:ssk1}. The kernel hyperparameters are estimated by marginal likelihood maximization, while $\sigma^2$ is estimated via least-squares residuals.
\item \textbf{KB-Or.}: Same as KB, with the difference that in this case the vector $z$ is available to the estimator. Hence, this estimator is regarded as an Oracle.
\item \textbf{ML-GS}: This is the method proposed in \cite{godoy2011identification}, based on maximum likelihood identification of the impulse response, namely
\begin{equation} \label{eq:MLFIR}
\hat g = \arg \max_{g,\,\sigma^2}  p(y;\,g,\,\sigma^2) \,.
\end{equation} 
To solve the likelihood problem, a scheme based on the EM method is proposed. In our implementation, the E-step of the EM iterations is computed using the Gibbs sampler (in contrast, \cite{godoy2011identification} proposes a scenario-based approach). The length of $g$ is also set to 50, while the convergence is established by using the same conditions as the estimator KB-GS-1.
\item \textbf{MAP-GS}: This is the method proposed in \cite{chen2012impulse}, based on maximum-a-posteriori (MAP) identification of the impulse response, namely
\begin{equation} \label{eq:MAPFIR}
\hat g = \arg \max_{g}  p(y|g;\,\sigma^2) p(g;\,\lambda,\,\beta) \,,
\end{equation} 
where the prior $p(g;\,\lambda,\,\beta)$ corresponds to \eqref{eq:model_gh}. Again, an EM solution scheme based on the Gibbs sampler is employed to solve this problem, setting $m=50$, while the convergence is established by using the same conditions as the estimator KB-GS-1. To facilitate hyperparameter and noise variance tuning, we plug those that are estimated by the method KB-Or., which has access to the non-quantized output $z$.
\end{itemize}
The performance of the estimators is evaluated by means of the fitting score, computed as
\begin{equation}
FIT_i  = 1-\frac{\|g^i - \hat g^i \|_2}{\|g^i\|_2} \,,
\end{equation}
where $g^i$ is the impulse response generated at the $i$-th Monte Carlo run and $\hat g^i$ the estimate computed by the tested methods.

\subsection{Binary quantizer}

The first experiment considers the following binary quantizer
$$
\mathcal Q_b [x] := \left\{ \begin{array}{ll} 1 & \mbox{ if } x \geq 1 \\ -1 & \mbox{ if } x < 1 \end{array} \right.\,.
$$
For each Monte Carlo run, the noise variance is such that $\frac{\var (Ug) }{\sigma^2} = 10$, i.e. the ratio between the variance of the noiseless (non-quantized) output and the noise is equal to 10. If $\sigma^2 > 2$, the system is discarded. We generate $N = 500$ data samples.
\begin{figure}[!ht]
\begin{center}
    {\includegraphics[width=\columnwidth]{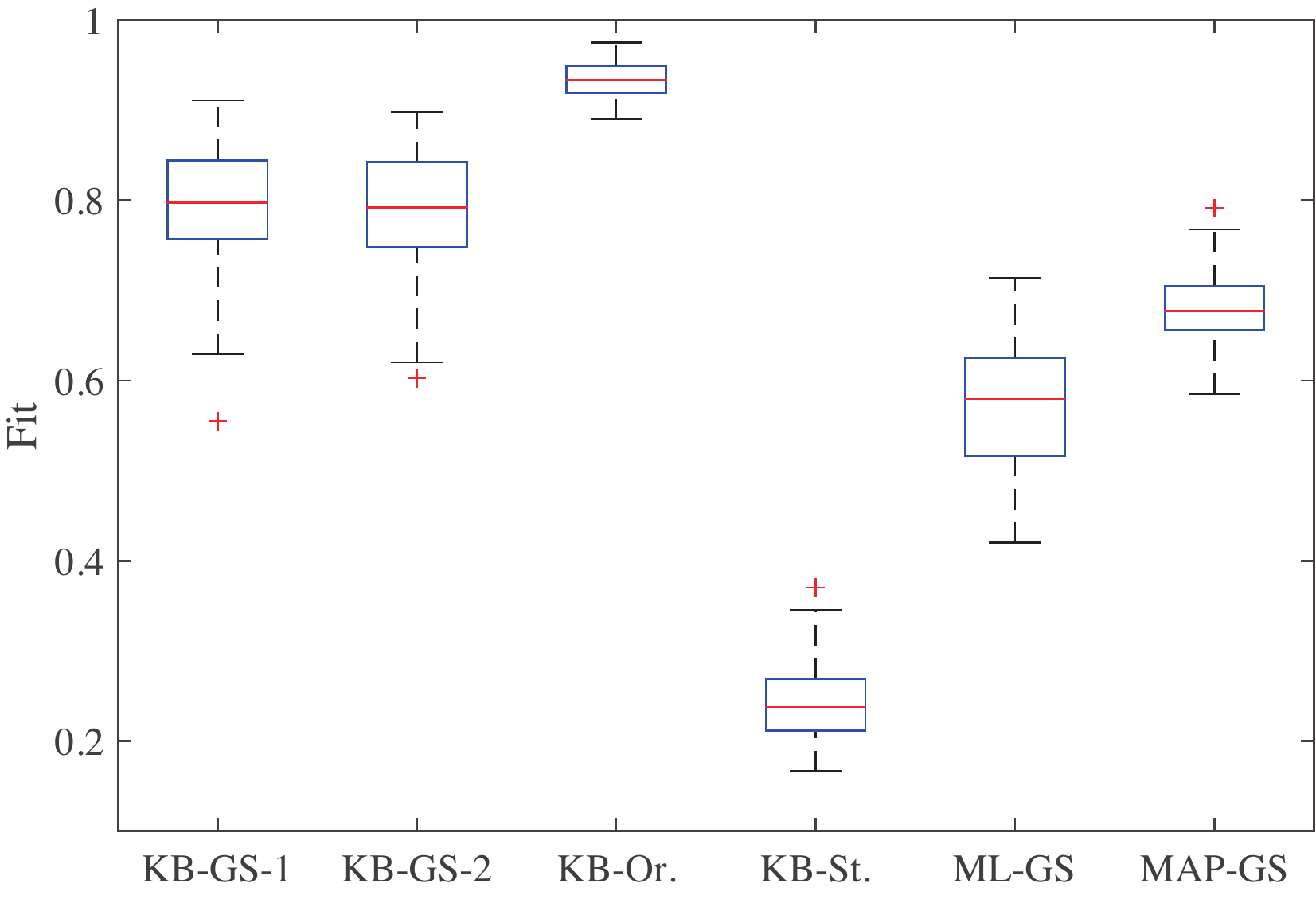}}
\caption{\emph{Box plots of the fitting scores for the binary quantizer experiment.}} \label{fig:bin_exp}
\end{center}
\end{figure}
Figure \ref{fig:bin_exp} shows the results of the Monte Carlo runs. The advantage of using the proposed identification scheme, compared to the method that does not account for the quantizer, is evident. Despite the large loss of information caused by the quantizer, the proposed method gives a fit which is quite comparable to the oracle method (KB-Or.). The proposed sampling mechanisms yield nearly equivalent performance. Furthermore, there seems to be a substantial advantage in using a Bayesian approach compared to the non-regularized estimator ML-GS, which is tailored for short FIR systems rather than IIR systems. The high number of coefficients to be estimated inevitably leads to high variance in the estimates. If this effect is not suitably alleviated by regularization (i.e., introducing a ``good'' bias), the performance of the estimator is doomed to be poor. Finally, the proposed estimators, based on computing the MMSE estimate of the impulse response, outperform the estimator MAP-GS, which is based on computing the MAP estimate of the impulse response.

\subsection{Ceil-type quantizer}

In the second experiment we test the performance of the proposed method on systems followed by a ceil-type quantizer, which is defined as
$$
\mathcal Q_c [x] := \lceil x \rceil \,.
$$
Again, for each Monte Carlo run, the noise variance is such that $\frac{\var (Ug) }{\sigma^2} = 10$. We generate $N = 200$ data samples.

\begin{figure}[!ht]
\begin{center}
    {\includegraphics[width=\columnwidth]{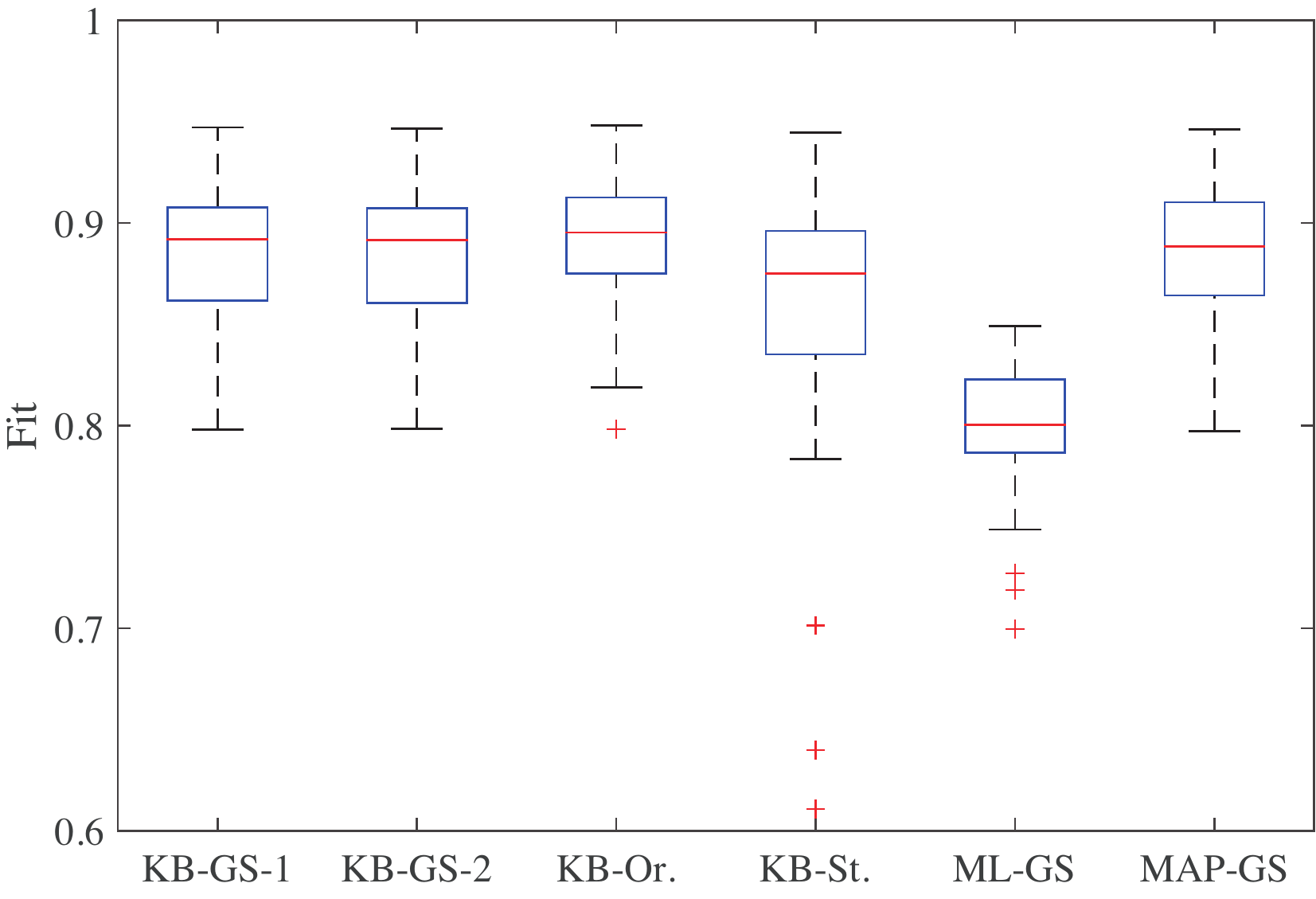}}
\caption{\emph{Box plots of the fitting scores for the ceil-type quantizer experiment.}} \label{fig:ceil_exp}
\end{center}
\end{figure}

As shown in Figure \ref{fig:ceil_exp}, in this case, if one compares the oracle-type method (i.e. KB-Or.) with the same method using quantized data (KB-St.), the loss of accuracy is relatively low. This is because this type of quantizer has a mild effect on the measurements. It can be seen, however, that the proposed methods are able to give a fit that is comparable to the oracle KB-Or., regardless of the employed sampling method. We notice also that the Bayesian approach has a major impact on the performance, as seen by the gap in the performance between ML-GS and the other estimators.

\section{Conclusions}
In this paper, we have introduced a novel method for system identification when the output is subject to quantization. We have proposed a MCMC scheme that exploits the Bayesian description of the unknown system. In particular, we have shown how to design two integration schemes based on the Gibbs sampler by exploiting the knowledge of the conditional probability density functions of the variables entering the system. The two sampling techniques can be used in combination with the EM method to perform empirical Bayes estimation of the kernel hyperparameters and the noise variance. We have highlighted, through some numerical experiments, the advantages of employing our method when quantizers affect the accuracy of measurements.

As a final remark, we note that the cascaded composition of a linear LTI dynamic system followed by a static nonlinear function is known in system identification as a \emph{Wiener system} \cite{giri2010block}, \cite{hagenblad2008maximum}. However, in Wiener systems the nonlinear function is generally assumed unknown, so a direct extension of the method proposed in this paper to general Wiener systems does not appear immediate, and would require the use of more involved and MCMC techniques (see, e.g., \cite{lindsten2013bayesian}). 

\bibliographystyle{plain}

\bibliography{biblio}

\end{document}